\documentclass[a4paper]{jpconf}
\usepackage{graphicx}
\usepackage{grffile}
\usepackage{amsmath, amsthm}
\usepackage{amsfonts,amssymb}
\usepackage{epstopdf}
\usepackage{pdfpages}
\pdfoutput=1
\begin{document}
%\includepdf[pages=1-12]{leghmouche_article.pdf}
\title{Effects of a quadrupolar magnetic term in a Generalized St\"{o}rmer problem}

\author{A Leghmouche and  N Mebarki}

\address{Laboratoire de Physique Math\'{e}matique et Physique Subatomique,\\ Fr\`{e}res Mentouri University of Constantine 1, Algeria}

\ead{aminaastro14@gmail.com}

\begin{abstract}
A new generalized St\"{o}rmer  problem is proposed. The charged particles motion around a rotating axisymmetric magnetic planet is studied using various conditions mainly in planetary magnetospheres. It is shown that the existence of specific trajectories is due to the quadrupolar and dipolar magnetic terms. Moreover, the equilibrium and stability behaviors and the corresponding orbital frequencies for various kinds of charged dust grains are treated giving rise to a new model describing the dynamical system and leading to more interesting results.
\end{abstract}

\section{Introduction}
Investigation of charged particles dynamics subject to the magnetic dipole has an important interest in astrophysics. In fact, their tiny fraction of mass in orbit around a planet \cite{1} was first introduced in the pioneering work of  St\"{o}rmer(1955)
\cite{2} mainly the theoretical study in planetary magnetospheres. It was also considered by Mendis and Axford(1974)\cite{3} in order to explain the origin of the accelerated mysterious dark spokes of Saturn's B ring in the space region observed by Voyager 1 and 2 \cite{4}.This treatment has been of a great interest to unveil the structure of force fields acting on the particles and their behaviors\cite{5}.It is worth to mention that the starting point to describe the motion of charged dust particles in a purely magnetic field orbiting  a planet  with a magnetosphere is the classical St\"{o}rmer problem \cite{6}.However, the consideration of the charged dust grain is constrained by the much small ratios between charge-to-mass which complicate the dynamics for the heaviest particles case. In the latter case, the planetary gravity and the co-rotational electric field have also to be taken into account and therefore the dynamics becomes more rich but complicated in the generalized St\"{o}rmer problem. In fact, this treatment is characterized by simplified assumptions of Keplerian gravity, specific magnetic field (pure dipole), constant charge without radiation pressure 
etc.. It is very important to remind that the four distinct St\"{o}rmer problems are:
 CSP(classical St\"{o}rmer problem) in which a charged particle moves in a pure dipole magnetic, RSP (rotational St\"{o}rmer problem) which is characterized by the electric field,GSP(gravitational St\"{o}rmer problem) in which only the Keplerian gravity is included and finally the most full interesting system RGSP(rotational gravitational St\"{o}rmer problem) in which both the fields action is taken into account 
\cite{7} \cite{8}.So far, many models giving exact analytical expressions have been developed.In fact, much work has been done on dynamics of charged dust grains near non spherical planets (prolate and oblate) \cite{9} and also near a black holes and compact stars under the effect of a strong magnetic field \cite{5}.Furthermore, the study of the stability of charged dust grains orbiting a planet characterized by all charge to mass ratios (from ions to rocks) \cite{10}  and sub-microns circumplanetary dust grains in multipolar fields case with the inclusion of non-axisymmetric magnetic field terms were investigated \cite{11}.In addition, the behavior of grains near solar corona with the influence of the solar radiation pressure\cite{12} and the injection of charged particles into a
magnetic dipole field (MDF) is also studied.\cite{13}

In the present work, we introduce the effect of a quadrupolar magnetic term (QMT) \cite{14} \cite{15}which is not included in the so called generalized St\"{o}rmer  problem.The main goal is to perform a detailed analytical treatment for axisymmetric magnetized planet by considering various parameters present in the effective potential \cite{16}.In  section 2 we presente the model, in section 3 we study the relative equilibrium with the effect of QMT.In section 4 the stability problem is discussed.Finally, in section 5 we draw our conclusion.
 Within our present model, we analyze the dynamics of charged dust particles verifying the magnetic dipole approximation \cite{7} and study their  relative equilibrium and stability in order to map the trajectories of the new dynamical system by giving specific Hamiltonian and effective potential. 
\section{The model description}
Our starting point is based on the approach of ref \cite{7}.We assume a
particle of mass $m$ and charge $q$  orbiting around a rotating
magnetic planet(aligned centered planet)\cite{16}  with mass $ M $ and
radius $R$,while the existence of both gravitational and
electromagnetic forces is crucial\cite{17}.\\
The general Hamiltonian of this particle is given by:
\begin{equation}
H=\frac{1}{2}m(\vec{P}-\frac{q}{c}\vec{A})^2 +U(r)
\end{equation}\\
Where $ c $ is the speed of light in the vacuum,$ \vec{r} $ the position vector of the particle,
 $ \vec{P}$  the conjugate momenta of   $ \vec{r} $  and  $ \vec{A} $  the vector potential with the following expression:
\begin{equation}
 \vec{A}(r)=\frac{\mu}{r^3}\left(
\begin{array}{c}
-y\\
x \\
0 \\
\end{array}
\right)
\end{equation}\\
Here $U(r)$ is the scalar potential which generates the electric and gravitational interactions.The magnetic field $ \vec{B}  $ of the 
planet is supposed to be a perfect magnetic dipole of strength $\mu$ aligned to the north-south poles of the planet.Otherwise,the magnetosphere surrounding the planet is taken as rigid conducting plasma that rotates with the same angular velocity $\vec{\Omega}
$ as the planet.
The charge ${q}$ is subject to a corotational electric field which has the general form:
\begin{equation}
 \vec{E}=-\frac{1}{c}(\vec{\Omega}\times \vec{r})\times\vec{B}
\end{equation}
In order to characterize the influence of the charge to the mass ratio of the charged particle, we assume that the gravitational and corotational electric effects depend on two parameters $\sigma_{g}$ and  $\sigma_{r}$ taking their values from 0 to 1\cite 7.In our treatment and in order to get more interesting results of the dynamic of charged particles near magnetic planet, we consider a general expression for the magnetic field taking into account the geomagnetic field \cite{18}. The expression reads (in the spherical coordinates system)
\begin{equation}
\overrightarrow{B}(r,\theta,\varphi,t)=-\overrightarrow{\nabla}
V(r,\theta,\varphi,t)
\end{equation} \\
 where   
 \begin{equation}
V(r,\theta,\varphi,t) =a \sum_{l=1}^{n}
\sum_{m=0}^{l}(\frac{a}{r})^{l+1}(g^{\l }_{m}(t)\cos(m\varphi)+h^{\l
}_{m}(t)\sin(m\varphi))P^{\l }_{m}(\cos\theta)
\end{equation}
and \begin{equation}
\overrightarrow{E}=-\frac{q}{c}\Omega\overrightarrow{\nabla}(\frac{a^3}{r}(-g^{0}_{1}\sin^{2}\theta)+(\frac{a^4}{2r^2}(3g^{0}_{2}\sin^{2}\theta\cos\theta))
\end{equation}
Here, $ r $, $ \theta$ and $\varphi$ are the radius, co-latitude and longitude in spherical coordinates, $ P^{\l }_{m}(\cos\theta)$ are the associated Legendre polynomials , $g^{\l }_{m}$ and  $h^{\l }_{m}$  represents the Gaussian coefficients describing the magnetic field and $a$  is the planet radius, $ l $ and $ m $ are spherical harmonic degree and order respectively \cite {19}.
It is very important to mention that the treatment used in ref \cite {7} is a special case of this expression($ a^3g^{0}_{1}=1$ and
$a^4g^{0}_{2}=0 $ ),where in our case the magnetic dipole and quadrupole terms are considered.In what follows ,we try to understand the effect of this QMT on the generalized St\"{o}rmer problem. 
%Here, $g^{\l }_{m}$ and  $h^{\l }_{m}$  represents the Gaussian coefficients depending on the time  and the Legendre polynomials  $ P^{\l }_{m}\cos\theta$   where $ l $  is the  quantum number of orbital angular momentum, and  $ m $ a magnetic number.The general formula of corotational electric field is used in our investigation with development depending on the terms with is :
%The parametric equation represents the corotational electric field
%using spherical harmonics, we can find the treatment \cite 6  as a
%special case of this expression($ a^2g^{0}_{1}=1$ and
%$a^3g^{0}_{2}=0 $ ), where besides a magnetic dipole another
%quadrupolar term is considered . The term proportional to is allowed
%in this study. the main question posed in this context is what is
%the results of the effect of this quadrupolar term on St?rmer
%problem for dynamics of charged dust grains  near magnetic planet.
In this paper one can show,the dynamics of charged dust grains is
described by the following Hamiltonian taking into account the effect of a QMT:
\begin{equation}
H=\frac{1}{2}(P^{2}_{r}+\frac{P^{2}_{\theta}}{r})+\frac{1}{2}{\omega}^2
r^2\sin^{2}\theta-\frac{ \sigma_{g}}{r}+\sigma_{r}\delta
\frac{a^3g^{0}_{1} \sin^{2}\theta}{r}+\sigma_{r}\delta \frac{3
a^4g^{0}_{2} \sin^{2}\theta\cos\theta}{2r^2}+ U_{eff}
\end{equation}
where the effective potential $U_{eff}$ has the expression \begin{equation}
U_{eff}=\frac{1}{2}{\omega}^2 r^2\sin^{2}\theta+\sigma_{r}\delta
\frac{a^3g^{0}_{1} \sin^{2}\theta}{r}+\sigma_{r}\delta \frac{3
a^4g^{0}_{2} \sin^{2}\theta\cos\theta}{2r^2}-\frac{ \sigma_{g}}{r}
\end{equation}
Here $\delta $ is the charge of the dust grain and  $\omega$ corresponds to the orbital
frequency.The notation $P_{r} $ and $P_{\theta} $ stand for the derivatives with respect to $ r $ and $ \theta $ respectively. To make our study clear and comprehensive and in order to compare our results concerning the trajectories of charged dust grains with those of ref \cite{7}, we take  $ a^3 g_1^0 $=$ a^4
g_2^0=1 $.
\section{Relative equilibrium  }
In order to study charged particles equilibrium,taking into account the QMT effect and 
find  the critical points,we have to solve the differential equations  system :
\begin{equation}
  \frac{\partial U_{eff} }{\partial \ r }  =0  
\end{equation}
and 
\begin{equation}
 \frac{\partial U_{eff} }{\partial \ \theta}  =0 
\end{equation} \\The sign of  the particle angular velocity  $\omega$ determines the
nature  of the charged dust grains motion( prograde or retrograde with a positive or
a negative charge respectively) with the direction of 
planet motion . The critical points of the  $U_{eff}$  are
found as the solutions of the system equations\\
%\begin{equation}
%\frac{\partial U_{eff} }{\partial \ r }=-{\omega}^2 r
%\sin^{2}\theta+\frac{ 1}{r} (\delta
%(\omega-\sigma_{r})\sin^{2}\theta+\sigma_{g})+\frac{1}{r^3}(3
%\delta(\omega-\sigma_{r})) \sin^{2}\theta\cos\theta=0
%\end{equation}
%\begin{equation}
%\frac{\partial U_{eff} }{\partial \ \theta}  =-{\omega}^2
%r^2\sin\theta\cos\theta-\frac{2\delta(\omega+\sigma_{r})\sin\theta\cos\theta
%}{r}+\frac{3
%\delta(\sigma_{r}-\omega)\sin\theta\cos^{2}\theta}{r^2}+\frac{3
%\delta(\omega-\sigma_{r})\sin^3\theta}{2r^2}=0
%\end{equation}

One can show that the equivalent non linear system Eqs(9),(10) can be simplified as :
\begin{equation}
-{\omega}^2 r
\sin^{2}\theta+\frac{1}{r}( \delta (
\omega-\sigma_{r})\sin^{2}\theta+\sigma_{g})+\frac{1}{r^3}(3\delta
 (\omega-\sigma_{r}) \sin^{2}\theta\cos\theta=0
\end{equation}
\begin{equation}
\frac{-\sin\theta\cos\theta}{r^2}({ \omega}^2{ \ r}^4+2
\delta(\omega-\sigma_{r})r+3 \delta
(\omega-\sigma_{r})(\cos\theta-\frac{\sin\theta\tan\theta}{2})=0
\end{equation}
It is worth to mention that the angular velocity $\omega$ is a function of  $\ r$  and $\theta$.
To obtain the trajectories of charged dust particles within this approach we follow the method used in ref \cite{7} using the generalized St\"{o}rmer problem.Furthermore, if  $ \theta=\frac{\pi}{2} $ (resp.$ \theta\neq\frac{\pi}{2} $) we get equatorial (resp.non equatorial or Halos) orbits,where their existence is strictly related to the values of  $\omega$  and $\delta$.
%relied on the values of  $\omega$  and $\delta$
%The system of non-linear equations  gives a part of the dipolar
%magnetic field influence parameters and our contribution by terms
%related to the $ a^3g^{0}_{2}$ ,looking at the system o it is
%evident that it depends on a radial and angular parameter $ r$ and
%$\theta $  for the both equations. Besides this to describe the
%%trajectories of charged dust particles in the new conditions we use
%the methods derived from the model of the generalized St\"{o}rmer
%%problem  \cite{6} Furthermore, if  $ \theta=\frac{\pi}{2} $ we get
%equatorial orbits, and when $ \theta\neq\frac{\pi}{2} $ we obtain
%the non-equatorial (Halo) orbits. where their existence is strictly
%relied on the values of  $\omega$  and $\delta$
\subsection{Equatorial orbits}
Assuming that the motion is in the equatorial plane, such kind of
orbits appear when $ \theta=\frac{\pi}{2} $,and Eqs(12) is satisfied leading to the parametric equation:
\begin{equation}
{\omega}^2{ \ r}^4+2 \delta(\omega-\sigma_{r})r+3
\delta
(\omega-\sigma_{r})(\cos\theta-\frac{sin\theta\tan\theta}{2})=0
\end{equation}

\begin{figure}[h]
\begin{minipage}{18pc}
\includegraphics[width=14pc]{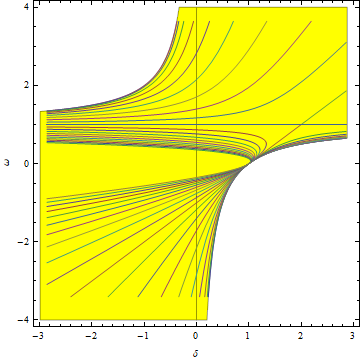}
\caption{\label{label}Equatorial orbits existence region for RGSP $
a^4g^{0}_{2}=1 $  }
\end{minipage}\hspace{2pc}%
\begin{minipage}{18pc}
\includegraphics[width=14pc]{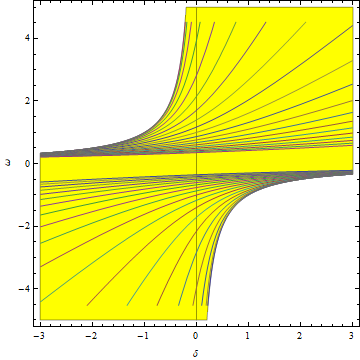}
\caption{\label{label}Equatorial orbits existence region for GSP $
a^4g^{0}_{2}=1 $}
\end{minipage}
\end{figure}
The motion of particles is well described in the 
$(\delta,\omega)$ plane for a generalized  St\"{o}rmer problem cases and
the presence of trajectories depends on the distance
from the center of the planet $r$ with retrograde and prograde
orbits.This situation is similar to the case $ a^3g^{0}_{1}=
a^4g^{0}_{2}=1 $ where the equatorial orbits exist
without the quadrupolar perturbations(the existence 
regions for all St\"{o}rmer problem generalizations are taken
into account).
\begin{figure}[h]
\begin{minipage}{18pc}
\includegraphics[width=14pc]{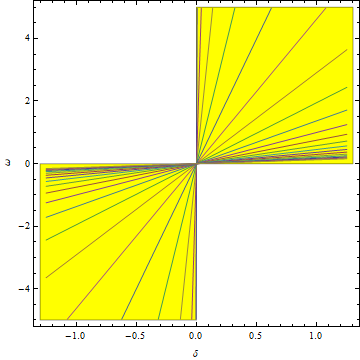}
\caption{\label{label}Equatorial orbits existence region for CSP $
a^4g^{0}_{2}=1 $}
\end{minipage}\hspace{2pc}%
\begin{minipage}{18pc}
\includegraphics[width=14pc]{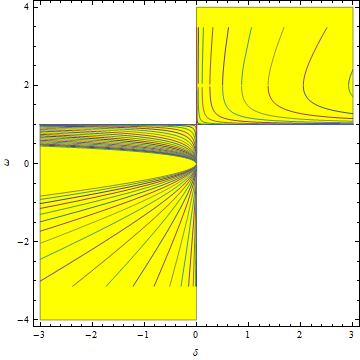}
\caption{\label{label}Equatorial orbits existence region for RSP $
a^4g^{0}_{2}=1 $}
\end{minipage}
\end{figure}
\subsection{Halos orbits}

Halos orbits ( do not cross the equatorial
plane) and appear when $ \theta\neq\frac{\pi}{2} $.There 
description is connected to the solution of the parametric equations
depending on the partial radial $ r $ and angular
$\theta$ parameters respectively.
They are  obtained  from the solution of Eqs (11),(12) system
%\begin{equation}
%Q(r , \omega)=0
%\end{equation}
%and
%\begin{equation}
%A(\theta , \omega)=0
%\end{equation}
%where 
\begin{eqnarray}
 Q(r ,\omega)&=&324{\delta }^4 {(\sigma_{r}-\omega)}^4+4{r}^4(-\sigma_{g}+\delta(\sigma_{r}-\omega) + {r}^3{\omega}^2)\nonumber\\&+&
(5{r}^3{\omega}^2 + \delta
(\omega-\sigma_{r})({{r}^3{\omega}^2} +{2 \delta\omega-\sigma_{r}}^2)
- (27{ \delta }^3{r}^2(\sigma_{r}-\omega)^2\nonumber\\&+& ((3
\sigma_{g} - 2 \delta \sigma_{r})^2 + 4 \delta (3 \sigma_{g} - 2 \delta
\sigma_{r})  \omega + 4 ( \delta^2 - 3\sigma_{g} {r}^3){\omega}^2 +
  8{ r}^6 {\omega}^4))=0
\end{eqnarray}
and
\begin{eqnarray}
 A(\theta,\omega)&=&\frac{27}{128}\delta {(\sigma_{r}-\omega)}^4 (7923 \delta^{3} \omega^{2} +1408\delta^{4}(\sigma_{r}-\omega)\cos\theta+13320{\delta }^3{\omega}^2 \cos 2\theta\nonumber\\&+&
576 \delta^{4} (\sigma_{r}-\omega)\cos 3\theta+( 7900
\delta^3{\omega}^2\cos( 4\theta) +(64\delta^{4}(\sigma_{r}-\omega)\cos5\theta)
\nonumber\\&-&(8\delta(\sigma_{g}(\sigma_{r}-\omega)(2\sigma_{g}\cos\theta({\sigma_{g}}^2+18(\delta^{2}{(\sigma_{r}-\omega)}^2\sin^{4}\theta+\sin\theta(8\delta{\sigma_{g}}^2\nonumber\\&&(\omega-\sigma_{r})\sin(2\theta))-({\sigma_{g}}^3+2\delta(\sigma_{r}-\omega)\sin^{2}\theta(-5{\sigma_{g}}^2
\nonumber\\&+&9\delta(\sigma_{r}-\omega)\sin^{2}\theta
(2\sigma_{g}+3\delta(-\sigma_{r}+\omega)\sin^{2}\theta)))\tan\theta
=0
\end{eqnarray}

To calculate halos orbits around a magnetic planet, the constraint of
the QMT is also considered, In this paper we
use analytical solutions to check the existence
of these trajectories %($d_{1}= a^2g^{0}_{1}\delta $  and $d_{2}=
%a^3g^{0}_{2}\delta $).
which gives Eqs(16) and (17) explicitly.
%deplacer sys eqts 
It is important to remark that the obtained family of curves  for the four known
generalized St\"{o}rmer cases with the inclusion of the
gravitational and co-rotational fields and the individual effect
are different from those of ref \cite 7.
\begin{figure}[h]
\begin{minipage}{18pc}
\includegraphics[width=18pc]{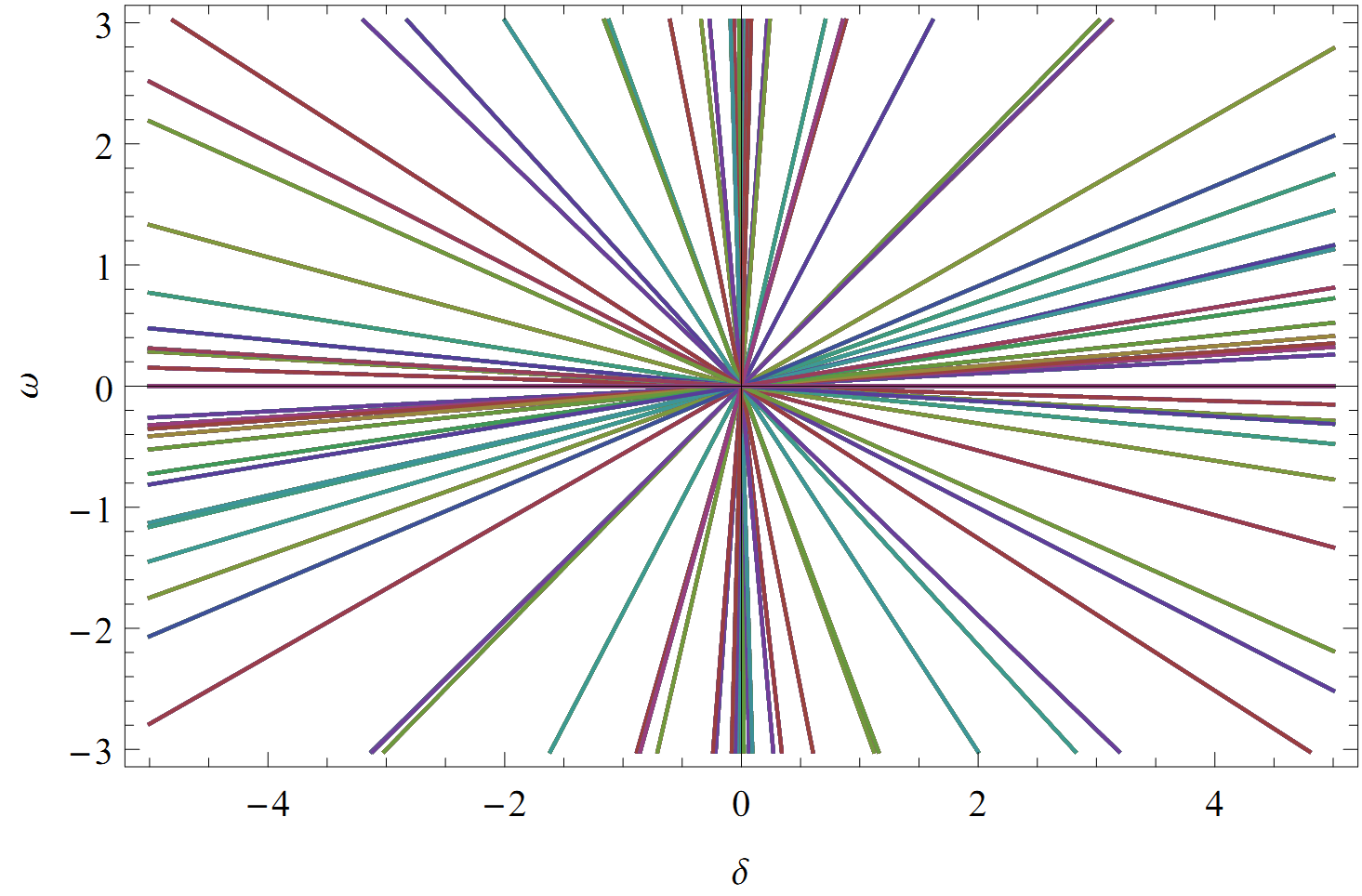}
\caption{\label{label}Curves of constant $r$ and $\theta$ for Halos
orbits (CSP ) with  $ a^4g^{0}_{2}=1$}
\end{minipage}\hspace{2pc}%
\begin{minipage}{18pc}
\includegraphics[width=18pc]{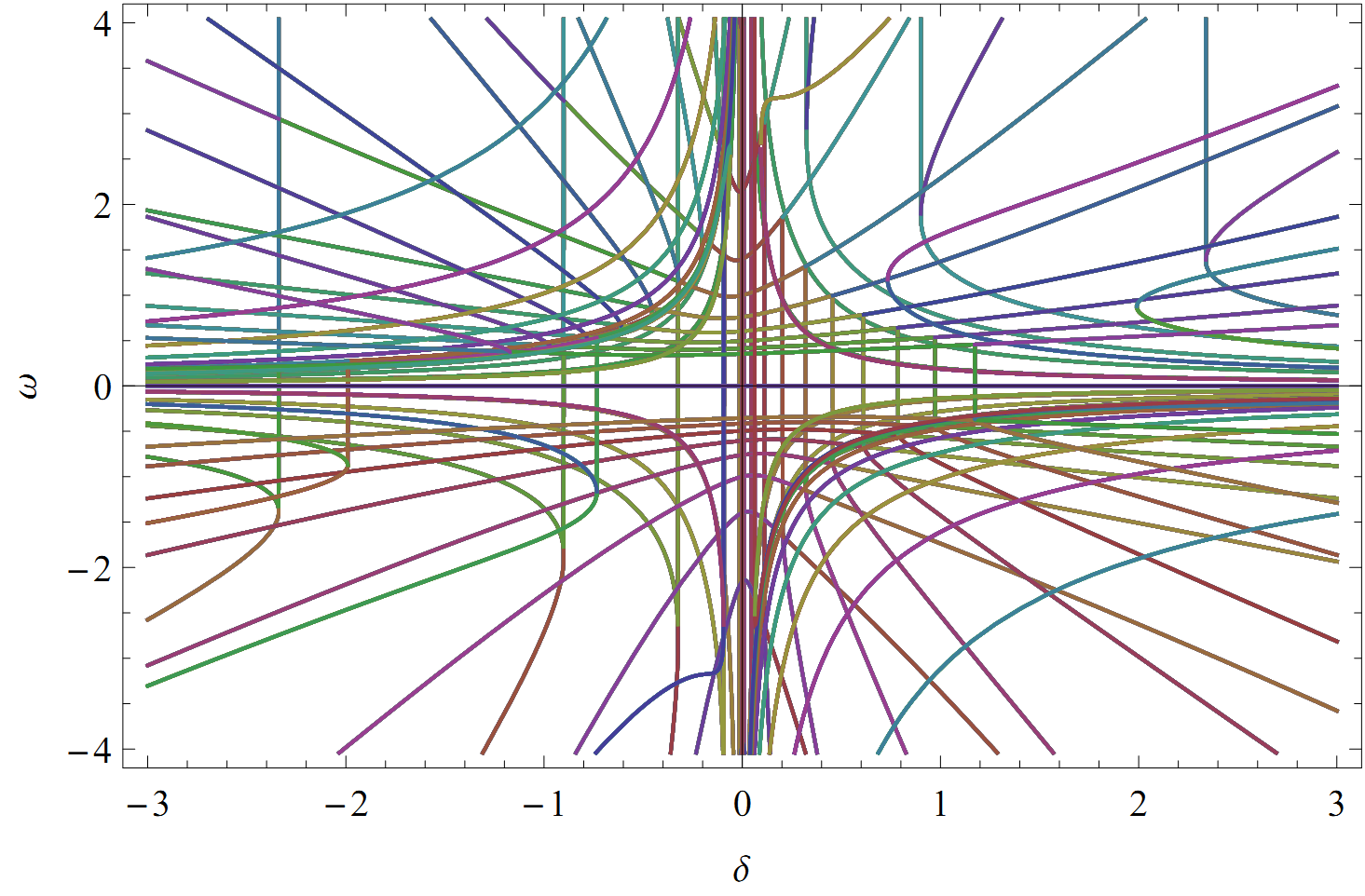}
\caption{\label{label}Curves of constant $r$ and $\theta$ for Halos
orbits (GSP ) with  $ a^4g^{0}_{2}=1$}
\end{minipage}
\end{figure}
\begin{figure}[h]
\begin{minipage}{18pc}
\includegraphics[width=18pc]{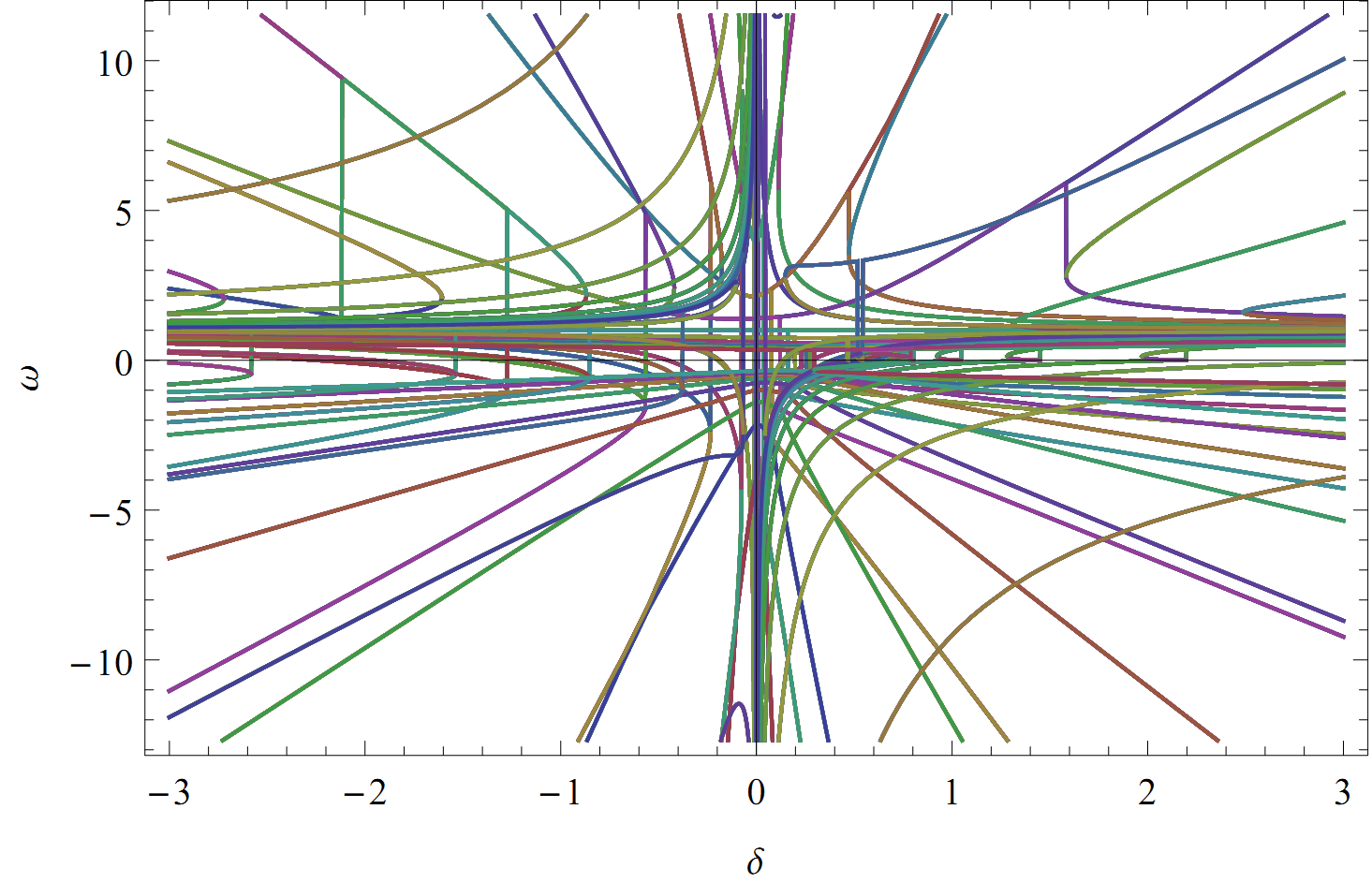}
\caption{\label{label}Curves of constant $r$ and $\theta$ for Halos
orbits (RGSP ) with  $ a^4g^{0}_{2}=1$}
\end{minipage}\hspace{2pc}%
\begin{minipage}{18pc}
\includegraphics[width=18pc]{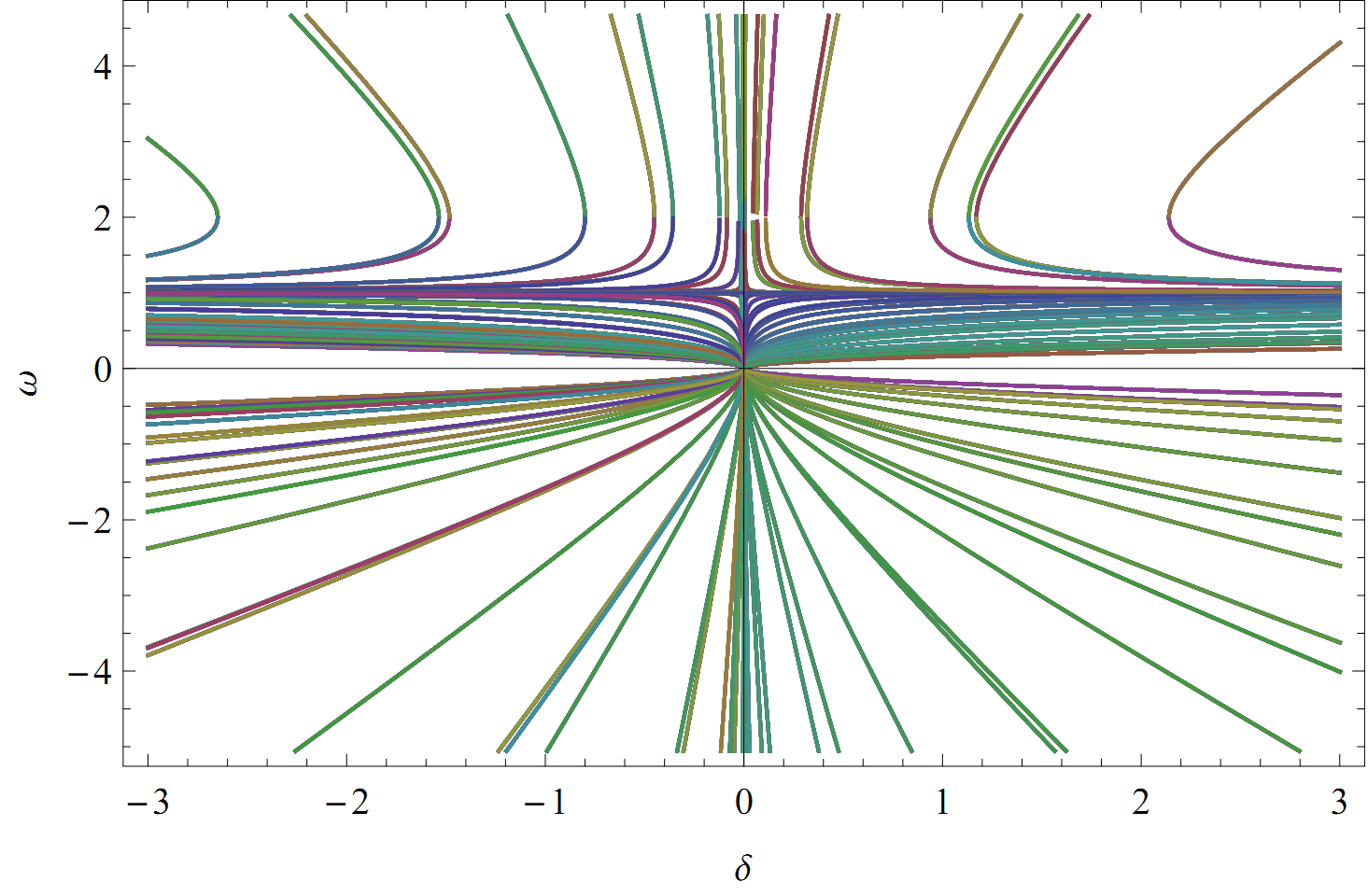}
\caption{\label{label}Curves of constant $r$ and $\theta$ for Halos
orbits (RSP ) with  $ a^4g^{0}_{2}=1$}
\end{minipage}
\end{figure}
%%%%%%%%%%%%%%%%%%%%%%%%%%%%%%%%%%%%%%%%%%%%%%%%

Our analysis has shown the possible families of curves defining
different particles trajectories related to both their nature and the planet rotation direction for the four St\"{o}rmer
problem. For the (CSP) (classical St\"{o}rmer
problem) in which a charged particle moves in a pure dipole magnetic
field, figure (5) displays in addition to the orbits found in ref \cite 7
they are others where the QMT is taken into account.In fact,it appears new orbits in regions which are not allowed in the model of ref \cite 7. Notice that, even in the absence of both co-rotational and gravitational fields,orbits exist for constant radius $r$  and angle $
\theta $. For the individual effect of the gravitational force, the
existence region is different from the St\"{o}rmer generalizations
where the orbits(get different behaviors  for  fixed radius values  $r$ and $ \theta $ (see figure
(6)).In the case of the presence of both the electric and magnetic
fields(adding the rotational and magnetic field (RGSP)), the
non-equatorial orbits exists in all regions for various charges.This is the most interesting case in the DMF
because it introduces simultaneous fields where the original case
is modified (and may be not circular orbits  for all cases see figure (7) ).
Figure (8) shows trajectories halos in the presence of the co-rotational electric field and absence of the gravitational field in this case  for fixed $ r $ and $ \theta $,negative as well as 
positives charges retrograde motion and prograde
motion for positive charges respectively. The used method in this work has the advantage to be clear and helps to describe
the profile of the trajectories and/or phase spaces more easily than the
numerical one.Our analysis has shown possible families  curves
defining various particles trajectories related to their nature and the planet rotation direction for the four
St\"{o}rmer problems cases.
\begin{figure}[h]
\begin{minipage}{18pc}
\includegraphics[width=18pc]{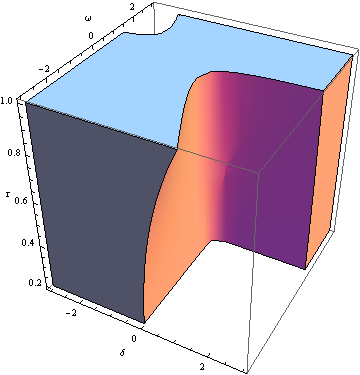}
\caption{\label{label}3D plot for RGSP with  $ a^4g^{0}_{2}=0$}
\end{minipage}\hspace{2pc}%
\begin{minipage}{18pc}
\includegraphics[width=18pc]{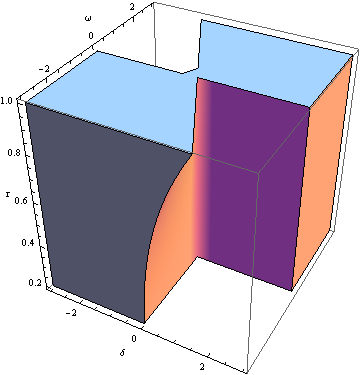}
\caption{\label{label}3D plot for GSP with  $ a^4g^{0}_{2}=0$}
\end{minipage}
\end{figure}
\begin{figure}[h]
\begin{minipage}{18pc}
\includegraphics[width=18pc]{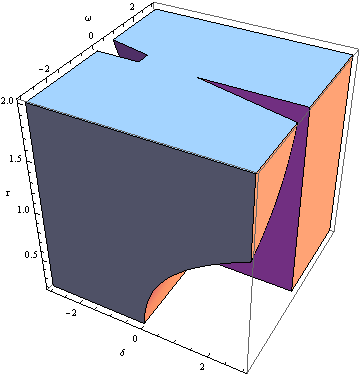}
\caption{\label{label}3D plot for CSP with  $ a^4g^{0}_{2}=0$}
\end{minipage}\hspace{2pc}%
\begin{minipage}{18pc}
\includegraphics[width=18pc]{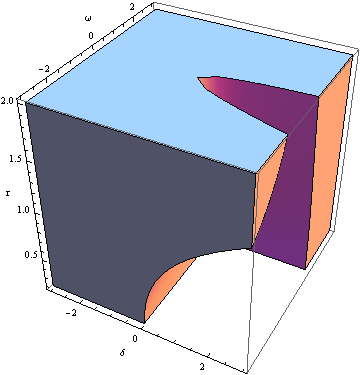}
\caption{\label{label}3D plot for RSP with  $ a^4g^{0}_{2}=0$}
\end{minipage}
\end{figure}

\subsection{Dynamical behavior and 3d plots}
Figures [9--16] (resp.[13--16]) show the dynamical behavior with DMF (resp.QMT) of radius $ r $ as a function of the orbital frequency $\omega$ and the charge  $\delta$.
Notice that,the individual and simultaneous effect of the gravitational and corotational electric fields modify the dynamics of charged dust grains particles located near a rotating magnetic planet.Furthermore,the introduction of a quadrupolar perturbations gives important results.

\begin{figure}[h]
\begin{minipage}{18pc}
\includegraphics[width=18pc]{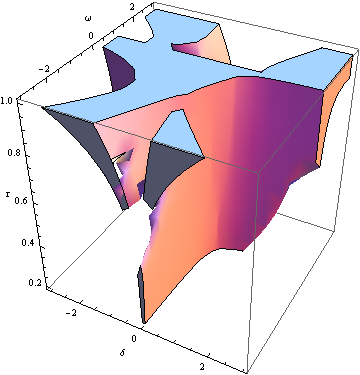}
\caption{\label{label}3D plot for RGSP with  $ a^4g^{0}_{2}=1$}
\end{minipage}\hspace{2pc}%
\begin{minipage}{18pc}
\includegraphics[width=18pc]{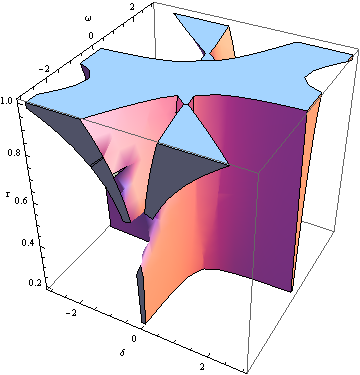}
\caption{\label{label}3D plot for GSP with  $ a^4g^{0}_{2}=1$}
\end{minipage}
\end{figure}

\begin{figure}[h]
\begin{minipage}{18pc}
\includegraphics[width=18pc]{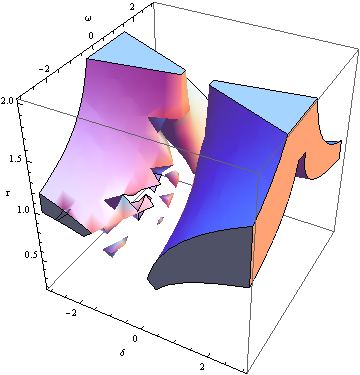}
\caption{\label{label}3D plot for CSP with  $ a^4g^{0}_{2}=1$}
\end{minipage}\hspace{2pc}%
\begin{minipage}{18pc}
\includegraphics[width=18pc]{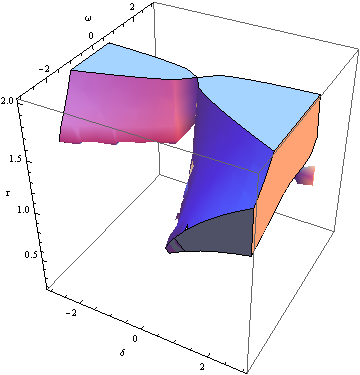}
\caption{\label{label}3D plot for RSP with  $ a^4g^{0}_{2}=1$}
\end{minipage}
\end{figure}
\begin{figure}[h]
\begin{minipage}{18pc}
\includegraphics[width=18pc]{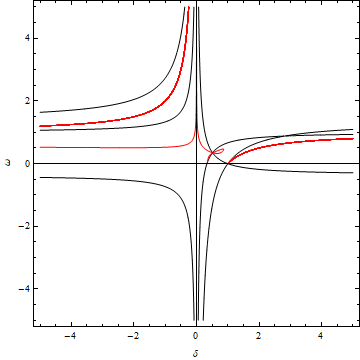}
\caption{\label{label}Stabilty regions for RGSP with  $
a^4g^{0}_{2}=1$}
\end{minipage}\hspace{2pc}%
\begin{minipage}{18pc}
\includegraphics[width=18pc]{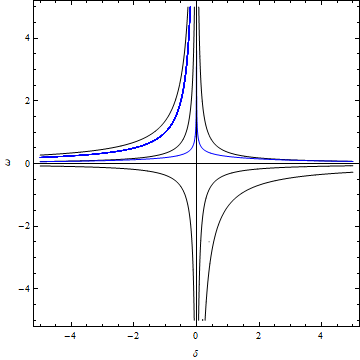}
\caption{\label{label}Stabilty regions for GSP $ a^4g^{0}_{2}=1$}
\end{minipage}
\end{figure}
\begin{figure}[h]
\begin{minipage}{18pc}
\includegraphics[width=18pc]{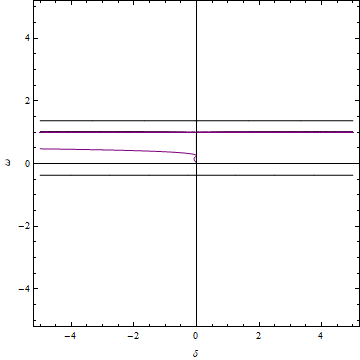}
\caption{\label{label}Stabilty regions for RSP with
$a^4g^{0}_{2}=1$}
\end{minipage}\hspace{2pc}%
\end{figure}
\section{Stability  }
In our treatment,stability of equatorial and halos orbits is related to $\delta$ and  
$\omega$ parameters.Following ref \cite{7},their stability is connected to the local minimum of the effective potential $U_{eff}$.The hessian matrix is defined by the second derivatives of $U_{eff}$ potential and given by:

\begin{equation}
\frac{\partial ^2U_{eff}}{\partial \ r^2}=\frac {
1}{r^{8}}\bigg(-5\sigma_{g}+\bigg( r^2(\delta^2+2\delta\
r^{3}(\sigma_{r}-3 \omega)+3\ r^6{\omega}^2)+3\delta
r(2\delta+r^{3}(3\sigma_{r}-7\omega))\cos\theta+9\delta^{2}\cos ^{2}
\theta\bigg)\sin^{2}\theta \bigg)
\end{equation}

\begin{eqnarray}
\frac{\partial ^2U_{eff} }{\partial \theta^2 }&=&\frac{
1}{32r^{6}}\bigg(99\delta^{2}+64 \delta^{2}r^{2}+128 \delta
r^{5}\omega+64 r ^{8}\omega^{2}+240\delta^{2}r \cos\theta-12 \delta
r^{4}\sigma_{r}\cos\theta+252 \delta r^{4}\omega
\cos\theta\nonumber\\&+&108 \delta^{2}\cos 2\theta+64
\delta^{2}r^{2}\cos 2\theta+64\delta r^{5}\sigma_{r}\cos 2\theta +
64\delta r^{5}\omega \cos 2 \theta+32 r^{8}\omega^{2}\cos 2
\theta\nonumber\\&+&144 \delta^{2}r \cos 3 \theta+108\delta
r^{4}\sigma_{r}\cos 3\theta+36\delta r^{4}\omega \cos
3\theta+81\delta^{2}\cos 4\theta\bigg)
\end{eqnarray}

\begin{eqnarray}
\frac{\partial ^2U_{eff}}{\partial r \partial \theta}&=&\frac {
1}{8 r^{8}}\bigg((\delta^{2}(387+112 r^{2})+32\delta
r^{5}(\sigma_{r}-3\omega)-48 r^{8}\omega^{2})\cos \theta+3\delta(76
\delta r+12 r^{4}\sigma_{r}\nonumber\\&+&16
r^{4}\omega+4r(29\delta+r^{3}(9\sigma_{r}-16\omega))\cos 2\theta+87
\delta\cos 3\theta)\bigg)\sin\theta
\end{eqnarray}
\begin{figure}[h]
\begin{minipage}{18pc}
\includegraphics[width=18pc]{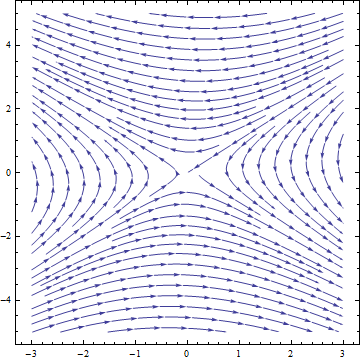}
\caption{\label{label}saddle unstable for RGSP }
\end{minipage}\hspace{2pc}%
\begin{minipage}{18pc}
\includegraphics[width=22pc]{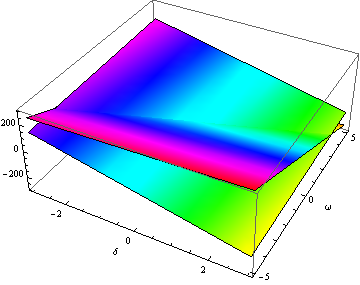}
\caption{\label{label}3d Curve of saddle unstable RGSP}
\end{minipage}
\end{figure}
\begin{figure}[h]
\begin{minipage}{18pc}
\includegraphics[width=18pc]{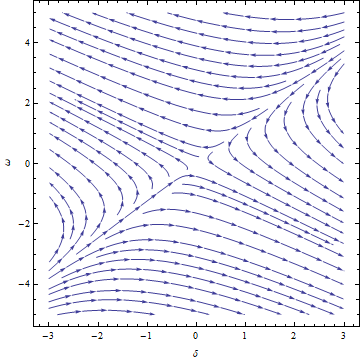}
\caption{\label{label}saddle unstable for GSP }
\end{minipage}\hspace{2pc}%
\begin{minipage}{18pc}
\includegraphics[width=22pc]{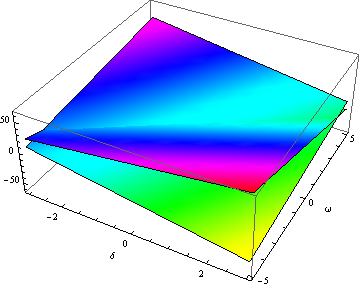}
\caption{\label{label}3d Curve of saddle unstable GSP}
\end{minipage}
\end{figure}
\begin{figure}[h]
\begin{minipage}{18pc}
\includegraphics[width=18pc]{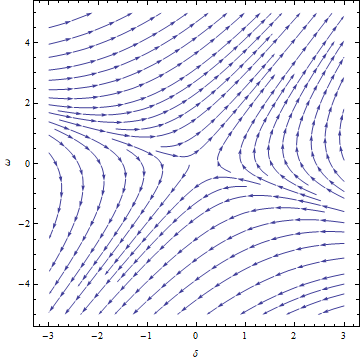}
\caption{\label{label}saddle unstable for RSP}
\end{minipage}\hspace{2pc}%
\begin{minipage}{18pc}
\includegraphics[width=22pc]{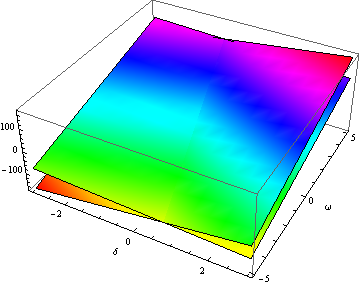}
\caption{\label{label}3d Curve of saddle unstable RSP}
\end{minipage}
\end{figure}
\begin{figure}[h]
\begin{minipage}{18pc}
\includegraphics[width=18pc]{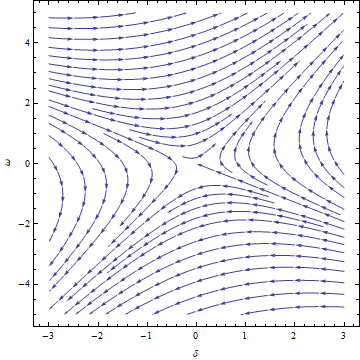}
\caption{\label{label}saddle unstable for CSP}
\end{minipage}\hspace{2pc}%
\begin{minipage}{18pc}
\includegraphics[width=22pc]{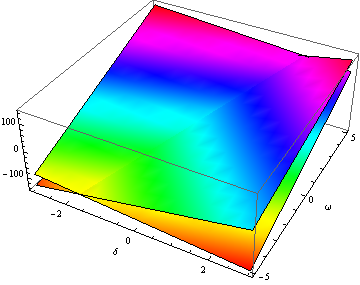}
\caption{\label{label}3d Curve of saddle unstable CSP}
\end{minipage}
\end{figure}
\subsection{Stability of equatorial orbits}
For the equatorial orbits ,we take $\theta =\frac{\pi}{2}$ and the 
radius $r$ from the generalization of kepler's third law
\cite{7}.Due to the obtained complicated equations where the mixed derivatives are not vanishing,the Hessian matrix determinant will allow to describe new stability regions in the equatorial plane (which does not exist in the work of ref\cite{7}) taking into account the QMT and for each case of a generalized St\"{o}rmer problem,the curves obtained from the Hessian matrix determinant are shown in figures [17--19].Notice that,for the two kinds of charged particles in various ranges of the orbital frequency $ \omega $.Significants modifications appear in the  region of stability treated with each effect of dipolar field.However,for the classical St\"{o}rmer case the trivial solutions can not explain the stability in these conditions.
\subsection{Stability of Halos orbits}
It is more complicated to define stability regions for this kind of trajectories.This is due to expression of the Hessian matrix with non vanishing partial derivatives.The substitution of the radial and angular parameters
$ r $ and  $ \theta $ leads to a relationship between $\omega$ and $\delta$
.In this paper,we have treated the fourth cases considered in ref \cite{7} with the individual and simultaneous effects of gravitational and corotational electric fields.Furthermore,to give a detailed physical description on the stability of Halos orbits  parallel to equatorial one needs  more computational sophisticated methods.Here,we have investigated the problem for
both negative and positive charged particles stability using boundary
conditions of the Hessian matrix and critical points.It is very important to point out that contrary to the dipolar approximation which present some stability regions (see figures [20--27]) the non equatorial orbits are unstable for positive and/or negative charged particles. 
%%%%%%%%%%%%%%%%%%%%%%%%%%%%%%%%%%%%%%%%%%%%%%%%
\section{Conclusion}
In this work we have studied new contributions compared to those of ref \cite{7} due to the quadrupolar component of the magnetic field with axisymmetric combination of gravitational and co-rotational electric fields for typical charge
to mass ratio.It turns out that,the QMT dynamics  perturbation give interesting results where the dynamics is more sensitive to the space parameters compared to the dipolar one and the occurrence of new orbits in the $\delta $- $\omega$ plane.Regarding the equilibrium state, the
existence of equatorial orbits of charged dust grains is mainly related to DMF  conditions  for
all St\"{o}rmer cases.For Halos(non-equatorial) orbits the
situation is more interesting,and their existence is significant with the simultaneous influence of co-rotational and gravitational
fields besides the cases where the individual effects of gravity
and electric field are not neglected giving rise to extra
trajectories which are not allowed in the regions presented in ref \cite{7}.Moreover,their behavior is more complicated leading to new interesting physical results.
Furthermore and contrary to ref \cite{7},small charge to mass constraint in all cases of fields
can give halos orbits at fixed $(r,\theta)$ parameters even if the gravity is not present. 
Notice that the stability of
trajectories is strongly dependent to the effect of $g _{2}^{0}$ .
Finally,our approach can be useful for the dynamics near spherical rotating magnetic planets by changing parameters $g _{m}^{l}$ in particular $g _{1}^{0}$ where in ref \cite{7} are taken to be $a^3g_{1}^{0}=1$ without any justification.Moreover it gives the nature of the motion of charged particles direction (prograde or retrograde) which can be very interesting from observational point of view taking into account not only the dipolar (like in ref  \cite{7} ) but the new quadrupolar contribution.We believe that other new results can be obtained for multipolar contributions (more study is under investigation) .

\section{Acknowledgments}
We are very grateful to the Algerian Ministry of Higher Educations
and Scientific Research and DGRSDT  for the financial support.

%%%%%%%%%%%%%%%%%%%%%%%%%%%%%%%%%%%%%%%%%%%%%%%%

\end{document}